\newcommand{\apj}{\textit{ApJ\ }}
\newcommand{\apjs}{\textit{ApJS\ }}

\newcommand{\mnras}{\textit{MNRAS\ }}
\newcommand{\aap}{\textit{A\&A\ }}
\newcommand{\nat}{\textit{Nature\ }}

\documentclass{iau}
\usepackage{graphicx,bm}

\title[Magnetic fields in early-type stars] 
{The nature and origin of magnetic fields in early-type stars}

\author[Jonathan Braithwaite]
{Jonathan Braithwaite}

\affiliation{Argelander Institut f\"ur Astronomie, Auf dem H\"ugel 71, 53121 Bonn, Germany \\ email: {\tt jonathan@astro.uni-bonn.de} \\[\affilskip]
}

\pubyear{2013}
\volume{302}  
\jname{Magnetic fields throughout stellar evolution}
\editors{P.\ Petit ... eds.}
\begin{document}

\maketitle

\begin{abstract}
I review our current knowledge of magnetic fields in stars more massive than around $1.5 M_\odot$, in particular their nature and origin. This includes the strong magnetic fields found in a subset of the population and the fossil field theory invoked to explain them; the subgauss fields detected in Vega and Sirius and their possible origin; and what we can infer about magnetic activity in massive stars and how it might be linked to subsurface convection.
\end{abstract}

\firstsection 
\section{Introduction}

Interest in stellar magnetism has increased in part because of the realisation that magnetic fields, together with rotation, are amongst the most important missing pieces in our understanding of how stars form, evolve and die. In star formation and various modes of stellar death such as $\gamma$-ray bursts magnetic fields are known to be crucial, and they also have subtler but important effects at other epochs in a star's life.

Here, I review magnetic fields which we can observe directly with the Zeeman effect, as well as those which have other observational signatures at the stellar surface.   [For an introduction to the Zeeman effect in stars and observational techniques to measure it, see Landstreet 2011.] This article does not include discussion of magnetic fields in stellar interiors which have no immediate effect at the surface, for instance small-scale fields which may be important in mixing and stellar evolution.

The focus here is early-type main-sequence (MS) stars. I begin by reviewing observations of the subset of early-type stars which display strong magnetic fields and discussing the theory of the fossil fields they are believed to contain. I then look at the rest of the population, first the intermediate-mass stars ($\lesssim7 M_\odot$) where subgauss fields have recently been detected and then the more massive stars. Finally, I discuss possible origins of magnetic fields and some open questions. The relevant observations and theories discussed here are summarised in Table \ref{tab1}. 

\begin{table}\begin{center}
  \caption{Summary of current knowledge of magnetic fields in early-type stars.}
  \label{tab1}
 \begin{tabular}{|l|c|c|}\hline 
 & {\bf A and late B} & {\bf O and early B} \\ \hline
{\bf Magnetic subset} & $B\sim200$ G to $30$ kG & $B\sim200$ G to $10$ kG \\
($\lesssim10$\%) & steady, large-scale & steady, large-scale \\
& Chemical peculiarities (Ap/Bp) & \\
& {\it Fossil field} & {\it Fossil field} \\ \hline
{\bf Rest of population} & Subgauss fields detected in two stars, & No direct detections \\
& probably present in all stars? & Indications of magnetic activity \\ 
& {\it Failed fossil field} & {\it Subsurface convection dynamo} \\ \hline
  \end{tabular}
 \end{center}\end{table}

\section{The magnetic subset of intermediate-mass stars: the Ap/Bp stars}\label{mag_int}

t is in these stars that the first magnetic fields were detected outside of the solar system (Babcock 1947), and not surprisingly it is these stars of which we have the best understanding in terms of magnetic properties. A subset of intermediate-mass stars are found to have strong magnetic fields which -- in contrast to the fields in convective stars -- are dominated by structure on large scales, and are not observed to vary in time. They range in strength from 200 G up to at least 30 kG. There is a real bimodality: whilst the chemically-peculiar Ap and Bp stars all have fields of at least about 200 G, no other intermediate-mass stars have fields above about a few gauss (Auri\`ere et al.\ 2007). 
 The magnetic fraction of the population depends on mass, starting at around 1\% at $1.5 M_\odot$ and rising to perhaps 10 or  20\% at $3 M_\odot$, where it seems to level off (Power et al.\ 2007).

\begin{figure}\begin{center}
\includegraphics[width=1.0\hsize,angle=0]{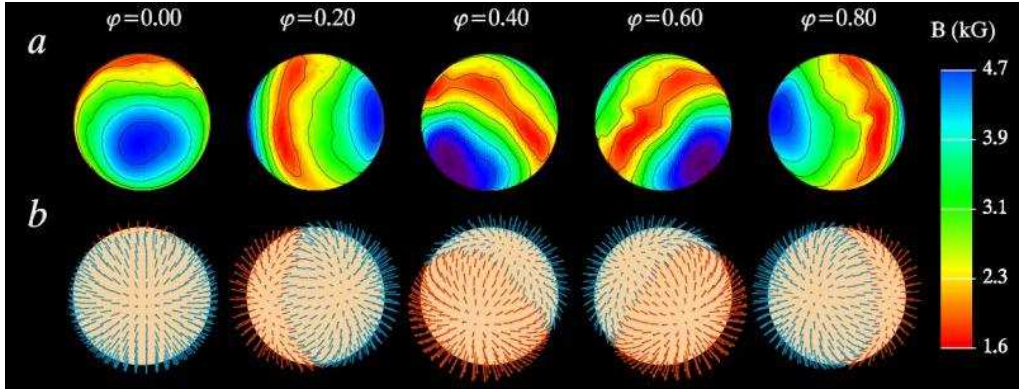}
\caption{The magnetic field of $\alpha^2$ CVn  viewed at five rotational phases. The top row shows field strengths and the bottow row the field direction. Clearly, $\alpha^2$ CVn has an approximately dipolar field which is inclined to the rotation axis. From Kochukhov et al.\ 2002. }
\label{a2cvn}
\end{center}\end{figure}

The magnetic fields display a range of geometries. A large fraction have fields to which a simple dipole is a good approximation, many have something which can be well described as an `offset dipole' or a dipole plus quadrupole or similar, and some have more complex fields which cannot be described adequately in terms of low-order spherical harmonics. An example of a star with a very simple field is shown in fig.\ \ref{a2cvn}.

As Cowling (1945) pointed out, the Ohmic dissipation timescale in the radiative core of the Sun is long, around $10^{10}$ yrs; consequently any magnetic field there finding itself in magnetohydrodynamic (MHD) equilibrium can in principle remain there for the Sun's entire lifetime. The same is true of more massive stars; according the the fossil field theory, magnetic stars with radiative envelopes contain just such an equilibrium.

There is not really any other plausible explanation for the strong, large-scale fields observed. For a long time it was discussed how the convective core may host a strong dynamo, the field produced there rising to the surface. However, whilst dynamo activity in the core is almost inevitable (e.g.\ Browning et al.\ 2004)
 it seems impossible to get the field to the surface on a sensible timescale (MacGregor \& Cassinelli 2003) and with the right geometry. In any case, similar fields to those seen in main-sequence stars have now been found in pre-MS stars without convective cores. Alecian et al.\ (2013) present the results from a survey of 70 Herbig Ae-Be stars, finding that the magnetic fraction is comparable to that amongst the main-sequence stars which these stars will become.

Historically the main challenge to the fossil field theory was to demonstrate that a stable MHD equilibrium can actually exist inside a star. With purely analytic methods it is relatively straightforward, given a star with a particular structure, to construct an equilibrium. Convincingly verifying its stability has proved impossible, however, although it has been possible to demonstrate that certain field configurations are {\it unstable}, for instance all axisymmetric fields which are either purely poloidal or purely toroidal (Wright 1973, Tayler 1973). The postulation of Wright (1973) and others that a stable configuration must therefore contain both components in a twisted-torus arrangement was confirmed by the discovery of exactly such configurations, using numerical methods (Braithwaite \& Spruit 2004).

Essentially, the method consists in evolving the MHD equations in a star containing initially some arbitrary field, as might be left over for instance by a convective dynamo or merger event (see section \ref{disc}). Braithwaite \& Spruit (2004) and Braithwaite \& Nordlund (2006) modelled a simplified radiative star: a self-gravitating ball of gas with an ideal gas equation of state, a polytropic index $n=3$ and ratio of specific heats $\gamma=5/3$ embedded in an atmosphere with low electrical conductivity. Over few Alfv\'en timescales, the field organises itself into an equilibrium, after which it continues to evolve only on the very long Ohmic timescale.

There is apparently a large range of equilibria available, and the particular equilibrium which appears depends on the initial conditions. In particular, it seems that the radial energy distribution of the inital field is important. Note that, during the process of relaxation to equilibrium, the gas in a radiative star/zone is restricted by gravity to move  around on spherical shells. Consequently it is impossible to transport flux in the radial direction so the total unsigned flux through any spherical shell $\oint |{\bf B}\cdot \rm{d} {\bf S}|$ can only fall. Therefore an initial field which is buried in the interior of a radiative star or zone evolves into a similarly buried equilibrium.

\begin{figure}\begin{center}
\includegraphics[width=0.35\hsize,angle=0]{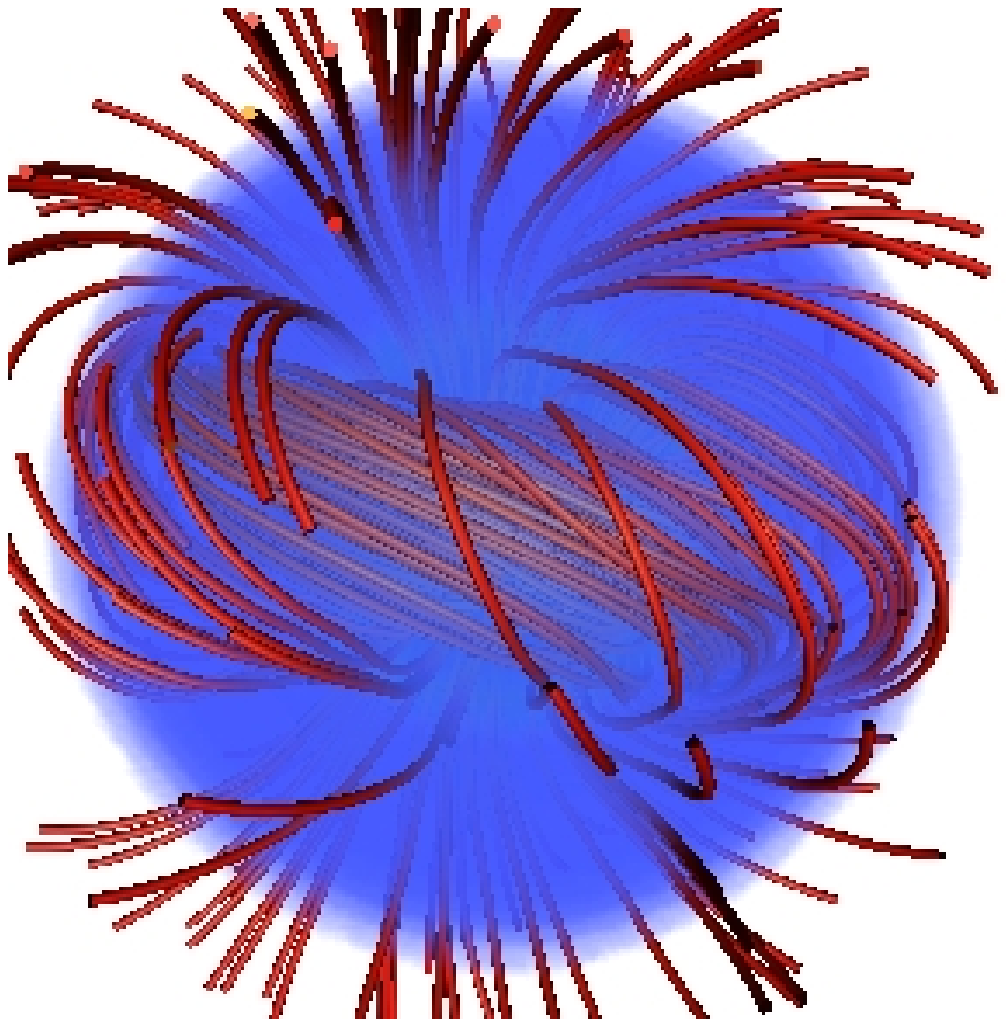}$\;\;\;\;$
\includegraphics[width=0.2\hsize,angle=0]{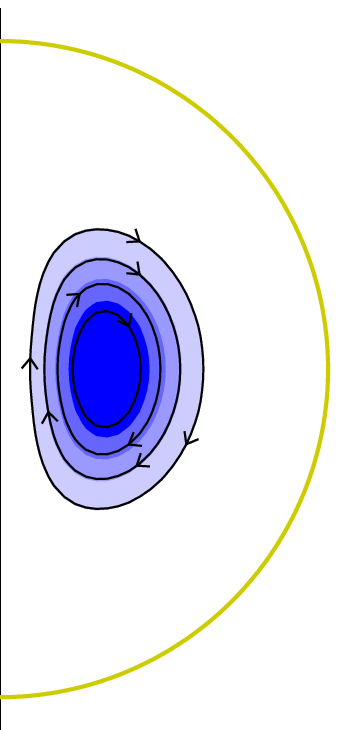}
\includegraphics[width=0.2\hsize,angle=0]{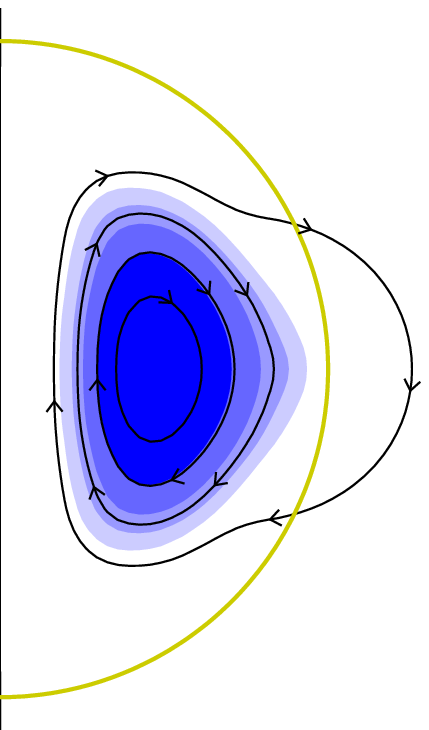}
\includegraphics[width=0.2\hsize,angle=0]{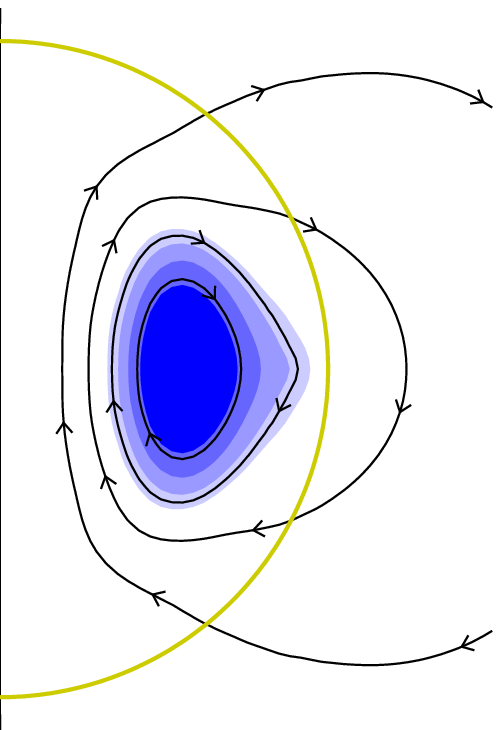}\\ \vspace{2mm}
\includegraphics[width=0.45\hsize,angle=0]{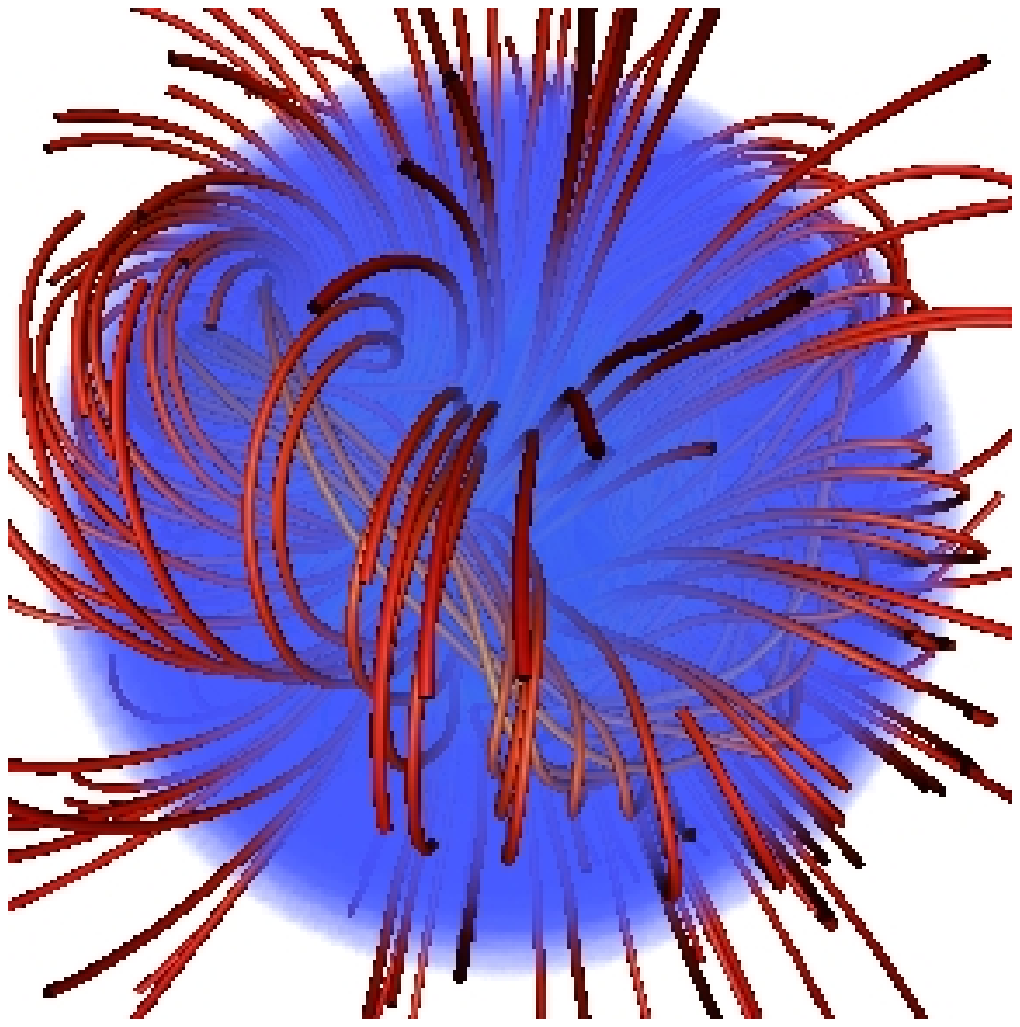}$\;\;\;$
\includegraphics[width=0.45\hsize,angle=0]{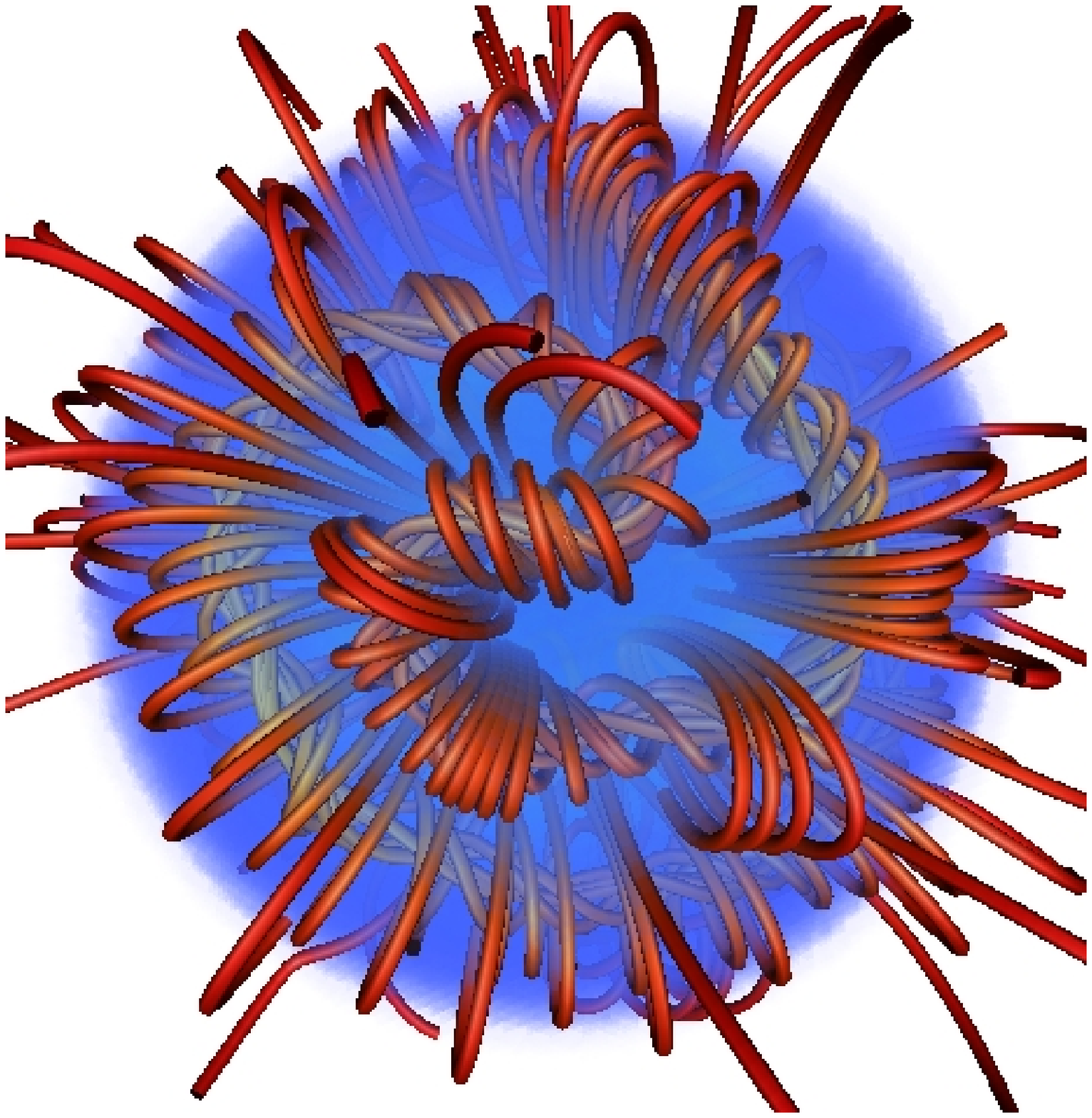}
\caption{Equilibria found in simulations. {\it Top left:} an axisymmetric twisted-torus field, consisting of a single twisted flux tube lying in a circle. {\it Top right:} cross sections of three such approximately axisymmetric equilibria, produced with different radial energy distributions in the initial conditions: on the left, the energy is most centrally concentrated; on the right, a flatter radial energy distribution. Blue shading represents the toroidal component of the field (multiplied by the cylindrical radius) and black lines are poloidal field. {\it Lower panels:} Non-axisymmetric equilibria: on the left, a somewhat flatter radial energy distribution than the equilibrium at the top right; on the right, a totally flat distribution. The latter corresponds qualitatively to those observed on stars such as $\tau$ Sco (see fig.\ \ref{tausco}). Figures from Braithwaite 2008 \& 2009.}\label{fig:simpics}
\end{center}\end{figure}

It turns out that if the initial field is somewhat stronger in the interior than near the surface, an approximately axisymmetric equilibrium evolves with both toroidal and poloidal components in a twisted-torus configuration, illustrated in fig.\ \ref{fig:simpics} (upper panels). This corresponds qualitatively to equilibria suggested by Prendergast (1956) and Wright (1973). If, on the other hand, the initial field has a flatter radial energy distribution and significant flux connects through the surface of the star, a more complex, non-axisymmetric equilibrium forms -- the lower panels of fig.\ \ref{fig:simpics}. It seems that both axisymmetric and non-axisymmetric equilibria do form in nature: see figs.\ \ref{a2cvn} and \ref{tausco}.

\begin{figure}
\parbox[t]{0.55\hsize}{\mbox{}\\ \includegraphics[width=1.0\hsize,angle=0]{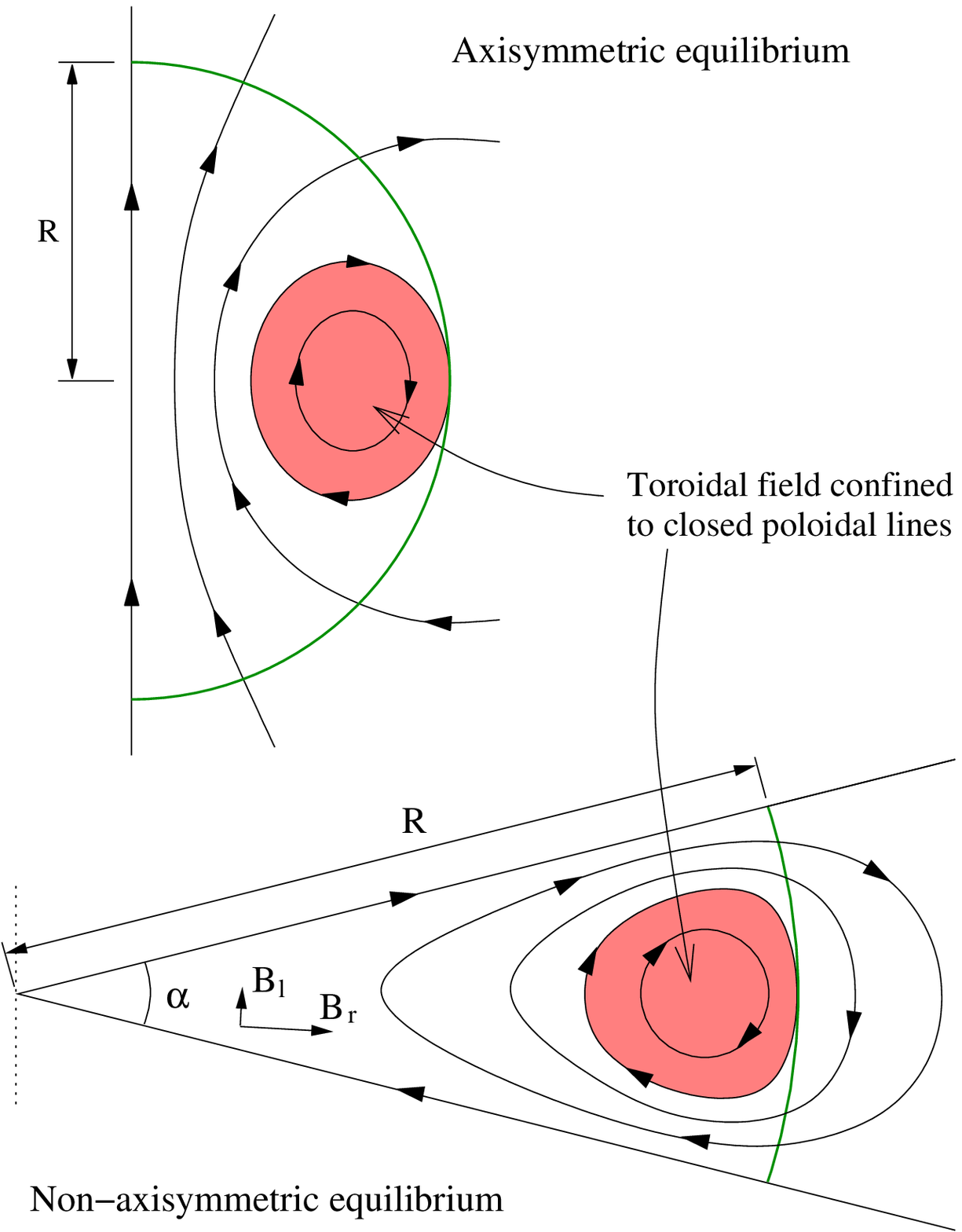}}
\hfill
\parbox[t]{0.4\hsize}{\vspace{1cm}\caption{Cross-sections of a twisted flux tube below the surface of a star. Above, the axisymmetric case where the flux tube lies in a circle around the magnetic equator; below, the non-axisymmetric case where the flux tube is narrower and meanders around the star in some apparently random fashion. The stellar surface is shown in green and poloidal field lines (in black) are marked with arrows. The toroidal field (direction into/out of the paper, red shaded area) is confined to the poloidal lines which are closed within the star. Toroidal field outside this area would unwind rather like a twisted elastic band not held at the ends. Figure from Braithwaite 2008.}
\label{fig:x-sec-diag}}
\end{figure}

The geometries of these various equilibria have one feature in particular in common, that they can be thought of in terms of twisted flux tubes (illustrated in fig.\ \ref{fig:x-sec-diag}). The simple axisymmetric equilibria can be thought of as a single twisted tube wrapped in a circle (a `twisted torus') and the more complex equilibria as one or more twisted flux tubes meandering around the star in apparently random patterns. In the equilibria found thus far, the meandering is done at roughly constant radius, a little below the surface. Equilibria where the flux tubes do not lie at constant radius seem possible but it also seems plausible that they are difficult to reach from realistic initial conditions, especially in view of the restriction of motion to spherical shells.

The properties of these non-axisymmetric and axisymmetric equilibria were explored further in Braithwaite 2008 \& 2009 respectively.

\section{The magnetic subset of high-mass stars}\label{mag_high}

\begin{figure}\begin{center}
\includegraphics[width=0.43\hsize,angle=0]{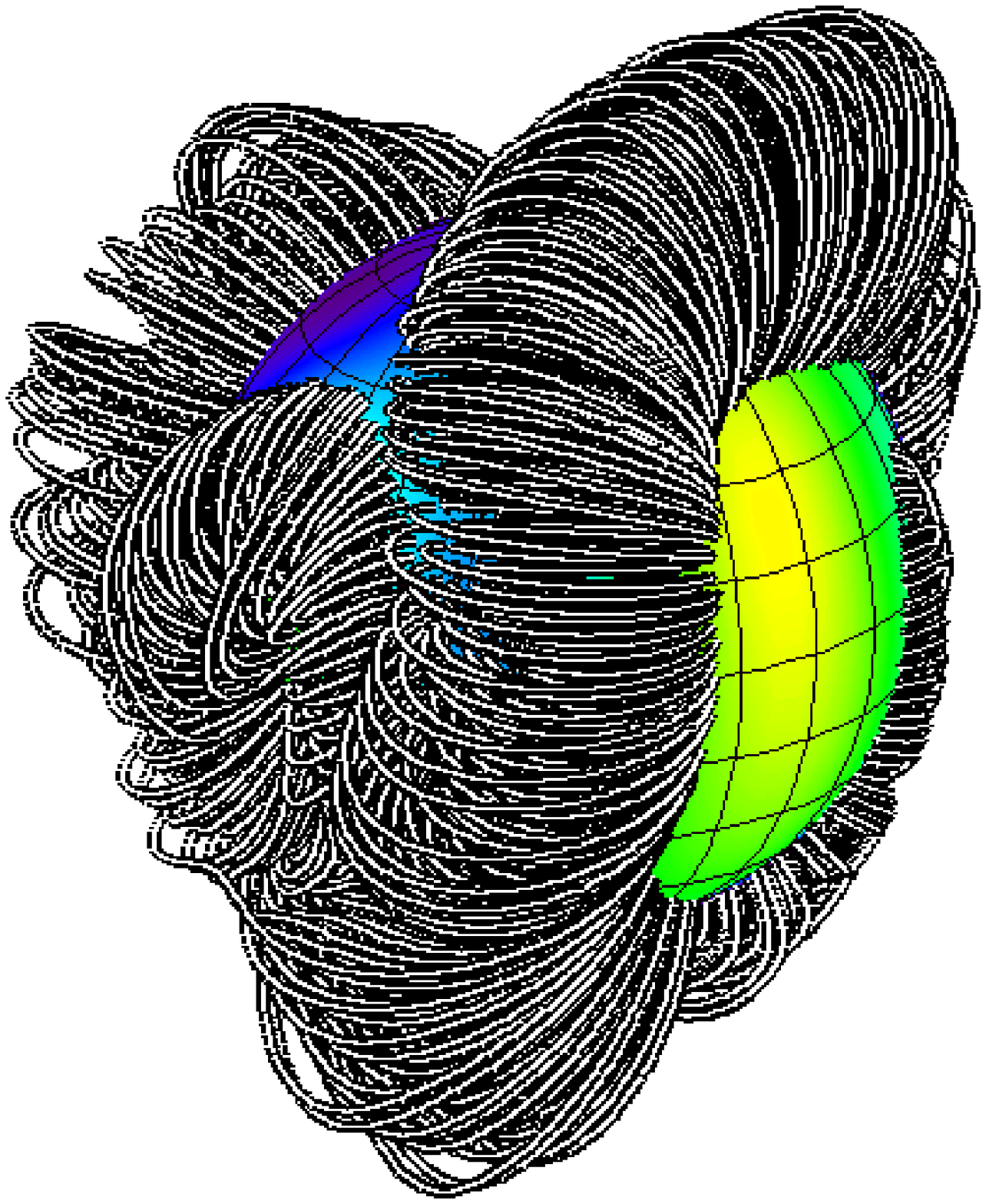}
\includegraphics[width=0.43\hsize,angle=0]{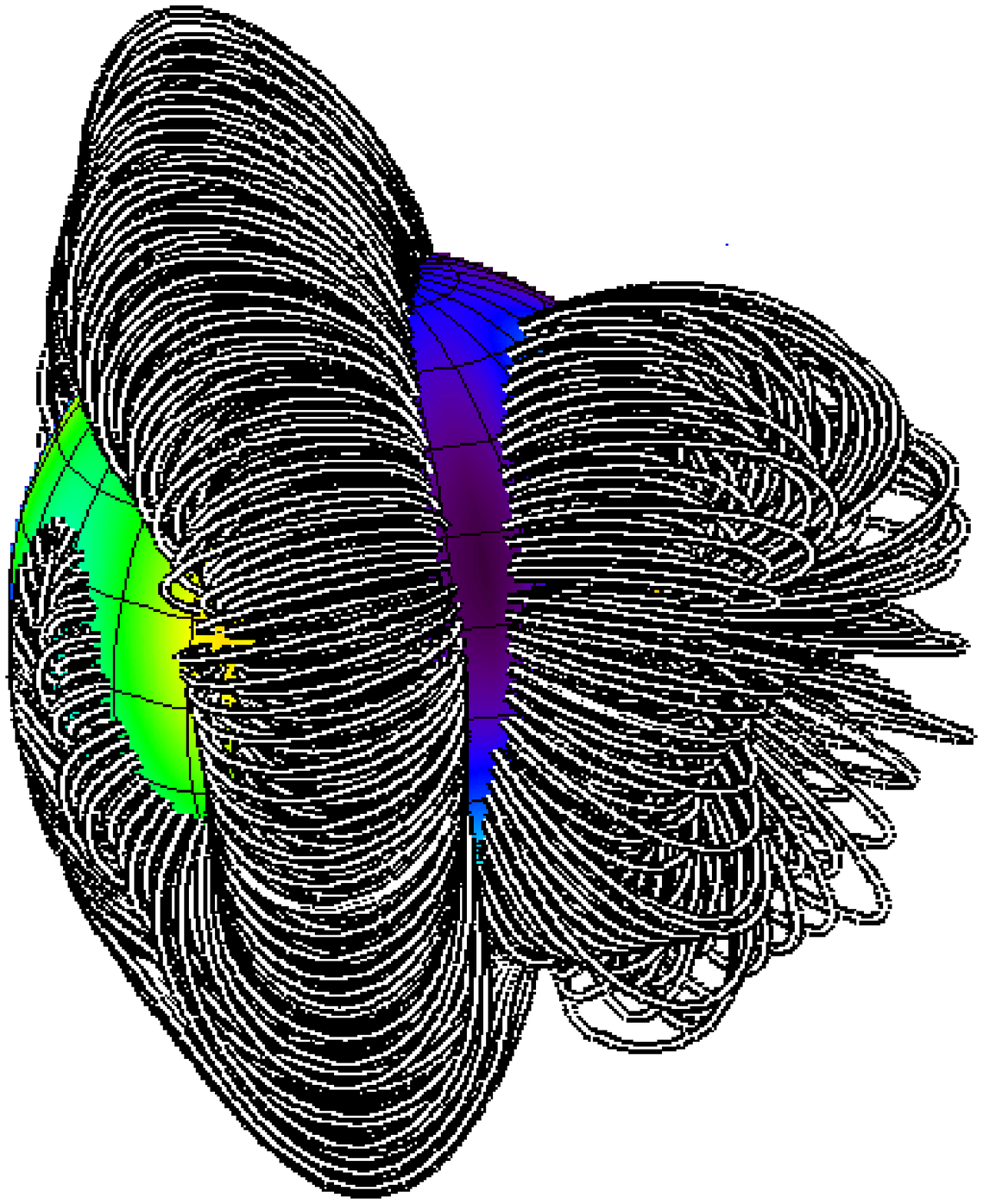}
\caption{The field geometry on the main-sequence B0 star $\tau$ Sco, at two rotational phases, using Zeeman-Doppler imaging. The paths of arched magnetic field lines are presumably associated with twisted flux tubes just below the surface. From Donati et al.\ 2006.}
\label{tausco}
\end{center}\end{figure}

For a multitude of reasons, the detection of magnetic fields in massive stars is more difficult than in lower-mass stars, but advances in technology and methods have improved the situation; see for Henrichs 2012 for a recent review. Thanks to recent surveys (e.g.\ Wade et al.\ 2013), we now know that around 10\% of the population host large-scale fields. The magnetic stars have fields of 200 G - 10 kG, a similar range to the intermediate-mass stars. Also just like the A stars, their fields have a variety of geometries: whilst some are approximately dipolar, others have a more complicated geometry -- see figure \ref{tausco} for an example. In several other stars similar magnetic fields have been found, dubbed the `$\tau$ Sco clones'.

It is tempting to conclude, therefore, that the phenomenon is simply a continuation of that seen in intermediate-mass stars, and indeed there are no particular theoretical reasons to think otherwise. The historical division between intermediate- and high-mass stars is probably due to difficulty observing the Zeeman effect in hotter stars, and that hotter magnetic stars do not develop chemical peculiarities. There may however be subtle differences due to the greater size of the convective core in hotter stars. If a fossil field is expelled somehow from the core, it might be difficult to maintain a perfect dipole shape at the surface and one expects something more complex; the surface field geometry for instance of $\tau$ Sco (figure \ref{tausco}) is consistent with a flat radial energy distribution with little or no magnetic field in the core (see section \ref{mag_int}).

\section{The rest of the intermediate-mass population}\label{nonmag_int}

As far as the `non-magnetic' intermediate-mass stars are concerned, there has recently been an exciting discovery of magnetic fields in Vega and Sirius, the two brightest A stars in the sky. Zeeman polarimetric observations have revealed weak magnetic fields: in Vega a field of $0.6Ê\pm 0.3$ÊG and in Sirius $0.2Ê\pm 0.1$ÊG (Ligni\`eres etÊ al.\ 2009, Petit etÊ al.\ 2011). The field geometry is poorly constrained, except that the field should be structured on reasonably large length scales, as cancellation effects would prevent detection of a very small-scale field. Also unconstrained is the existence or otherwise of time variability Ð although Petit et Êal.\ note that in Vega `no significant variability in the field structure is observed over a time span of one year'. Given that we have detections in both stars observed, it seems very likely that the rest of the `non-magnetic' population also have magnetic fields of this kind. There are also theoretical grounds to expect this.

 The only model of these fields we have so far is the so-called failed fossil hypothesis (Braithwaite \& Cantiello 2013). To understand this model, it is necessary first to consider the formation of the fossil fields described in the previous section. It is informative to compare sizes of terms from the momentum equation
\begin{eqnarray}\label{eq:momentum}
\frac{{\rm d}{\bf u}}{{\rm d} t} \;\;= \;\;-\frac{1}{\rho}{\bm \nabla} P \;\;+\;\; {\bf g}_{\rm eff}\;\; + \;\;\frac{1}{4\pi\rho}\bm{\nabla}\times {\bf B}\times{\bf B}\;\;-\;\;2{\bf \Omega}\times{\bf u}\\
\nonumber U^2 \,\;\;\;\;\;\;\;\;\;\;\;\;\;\; c_{\rm s}^2 \;\;\;\;\;\;\;\;\;\;\; v_{\rm ff}^2 \;\;\;\;\;\;\;\;\;\;\;\;\;\;\;\;\;\;\;\; v_{\rm A}^2 \;\;\;\;\;\;\;\;\;\;\;\;\;\;\;\;\;\;\;\; R\Omega U \;\;
\end{eqnarray}
where $U$, $c_{\rm s}$, $v_{\rm ff}$, $v_{\rm A}$, $R$ and $\Omega$ are the typical flow speed, sound speed, free-fall velocity, Alfv\'en speed, length scale (comparable to the stellar radius), and angular frequency of the star's rotation. In the radial direction, obviously the pressure gradient and gravity are in almost perfect balance (note that ${\bf g}_{\rm eff}$ includes the centrifugal force). In directions normal to gravity, i.e.\ on spherical shells, the magnetic field gives rise to motions. In the slowly rotating case where $R \Omega \ll v_{\rm A}$, the Coriolis force is small and the Lorentz force is balanced by inertia; consequently the flow speed $U$ is comparable to $v_{\rm A}$, and the magnetic field evolves on the timescale $R/U \sim \tau_{\rm A}$ where $\tau_{\rm A}\equiv R/v_{\rm A}$ is the Alfv\'en timescale. As an equilibrium is approached, inertia dies away and its role in balancing the Lorentz force is taken over by the pressure gradient and gravity, which is made possible by non-spherical adjustments to the pressure and density fields.

In a quickly-rotating star where $R \Omega \gg v_{\rm A}$ (accounting for almost all stars), the inertia term is small and the Lorentz force is balanced instead (on spherical shells) by the Coriolis force. This means that the flow speed $U\sim v_{\rm A}^2/R\Omega$ and the field now evolves on the timescale $\tau_{\rm A}^2\Omega$ instead of $\tau_{\rm A}$.

According to the failed fossil theory the magnetic field, instead of having reached an equilibrium long ago (as in the strongly magnetic stars), is still evolving towards equilibrium. Quantitatively, equating the age of the star to the evolution timescale $\tau_{\rm A}^2\Omega$, one can work backwards to find the field strength. In the cases of Vega and Sirius this gives $15$ and $5$ gauss respectively, the difference coming mainly from Vega's faster rotation. In other words, a field of $15$ G evolves dynamically in Vega on a timescale of $400$ Myr; it is impossible for a strong field left over from e.g. a pre-MS convective dynamo to decay below that strength in the time available. This somewhat surprising result has an analogy in terrestrial weather systems, where inertia is rarely important and pressure differences would be equalised on the sound-crossing time if the Earth were not rotating.

It is easy to reconcile these predicted field strengths with observed field strengths of $0.6$ and $0.2$ G: the observations will underestimate the strength of a smaller-scale field, and one naturally expects the surface field to be weaker than the predicted volume-average. Finally since we expect that all of these stars hosted a convective dynamo during the pre-MS, this theory predicts that {\it all} intermediate-mass stars without strong fossil fields should have magnetic fields of this kind. Finally, note that strongly magnetised stars will still have reached their equilibria relatively quickly: for a field strength of $3$ kG and a rotation period of $10$ days, $\tau_{\rm A}^2\Omega\sim 2000$ yr.

Note that the failed fossil theory does not take account of various processes which may be going on during the magnetic field's attempt to reach an equilibrium, such as meridional circulation and associated differential rotation; these process may tend to increase the field strength. Further work is necessary to study these effects.

\section{The rest of the high-mass population}\label{nonmag_mass}

There have been no direct Zeeman detections of magnetic fields in the rest of the high-mass population, and given that the detection limits are generally somewhat lower than the fields detected so far, it is probably the case that massive stars have the same magnetic bimodality as intermediate-mass stars. All high-mass stars, however, display a wealth of activity and variability not seen in the intermediate-mass stars: line profile variability, discrete absorption components, wind clumping, solar-like oscillations, red noise, photometric variability and X-ray emission (see e.g.\ Oskinova et al. 2012 for a review of these phenomena). Undoubtedly, these are at least partly caused by the strong radiation-driven winds and by the line-deshadowing instability which these winds are subject to, which can produce shocks. However, it is difficult to explain many of these phenomena without also invoking some kind of magnetic activity at the surface.

In principle, massive stars could contain failed fossil fields as described above. Indeed, the younger ages give even stronger fields, although an important difference may be that massive stars never pass through a pre-MS convective phase and begin the MS already with very weak fields. In any case, the presence of convection in layers just below the surface probably makes this irrelevant, as these are expected to produce fields which are stronger and have more interesting observational consequences. 

\begin{figure}[b]
\parbox[t]{0.4\hsize}{\vspace{1cm}\caption{Schematic of the magnetic field generated by dynamo action in a subsurface convection zone. Note that magnetic features should appear as bright spots on the surface, rather than as dark spots as in stars with convective envelopes. From Cantiello \& Braithwaite 2011.}
\label{fig:subsurf}}
\hfill \parbox[t]{0.54\hsize}{\mbox{}\\ \includegraphics[width=1.0\hsize]{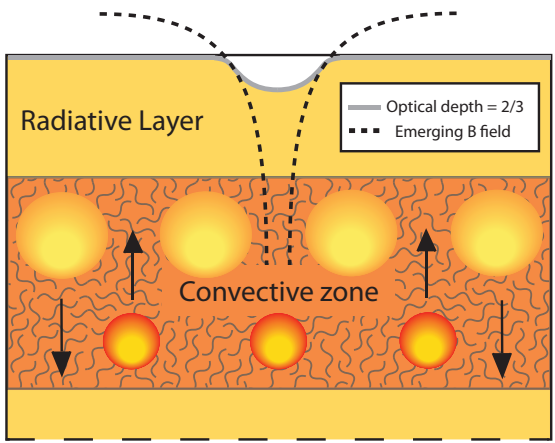}}
\end{figure}

As described by Cantiello et al.\ (2009), massive stars contain two or three thin convective layers close to the surface. These arise because of bumps in the opacity at certain temperatures, which in turn are caused by the ionisation of iron and helium. They are likely to host dynamo activity. Cantiello \& Braithwaite (2011) proposed that a magnetic field thus produced can easily rise buoyantly to the surface, thanks to the low density and consequently very short thermal timescale in the overlying layer, which allows heat to diffuse rapidly into magnetic features. This is illustrated in fig.\ \ref{fig:subsurf}. Note that magnetic pressure causes the photosphere to be lower inside magnetic features than in the surroundings. Consequently, magnetic spots on massive stars look bright. This contrasts with spots on convective stars, where the magnetic field has the additional effect of inhibiting upwards heat transfer.

\begin{figure}\begin{center}
\includegraphics[width=0.95\hsize,angle=0]{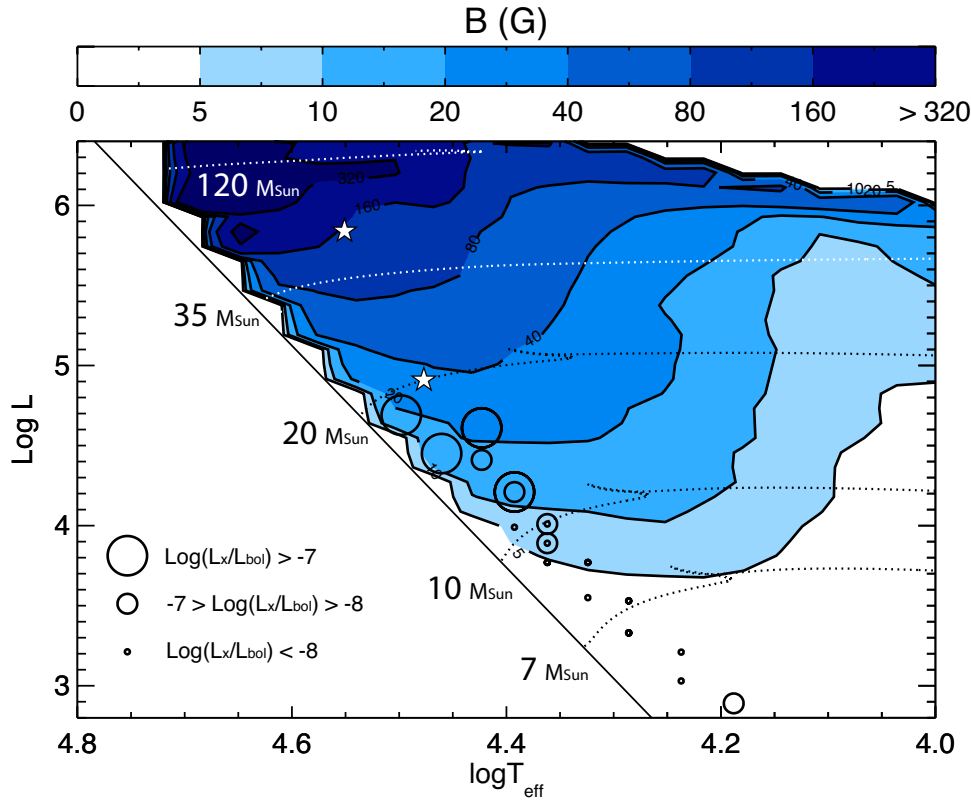}
\caption{The field strengths predicted at the surfaces of massive stars. This assumes an equipartition dynamo in the convective layer and a $B\propto\rho^{2/3}$ dependence as the field rises through the overlying radiative layer. Also shown are the X-ray luminosities measured in a number of stars, which is at least suggestive of a connection between subsurface convection and X-ray emission. From Cantiello \& Braithwaite 2011.}
\label{fig:bsur}
\end{center}\end{figure}

The deepest of these layers -- that associated with iron ionisation -- is energetically the most interesting. Assuming (a) an equipartition dynamo with $B^2/8\pi\sim\rho u^2/2$ and (b) that $B\propto\rho^{2/3}$ during the buoyant rise of magnetic features through the radiative layer, corresponding to isotropic expansion, field strengths of approximately $5$ to $300$ G are predicted -- as depicted in fig.\ \ref{fig:bsur} (for solar metallicity). The field strength depends on the mass and age of the star: higher fields in more massive stars and towards the end of the main sequence. These fields are expected to dissipate energy above the stellar surface and could give rise to, or at least play some role in, the various observational phenomena which are absent in the intermediate-mass stars. Indeed, if the X-ray luminosities of various main-sequence stars are plotted on the HR diagram (fig.\ \ref{fig:bsur}) a connection with subsurface convection does seem apparent. Of course, we cannot be certain that it is a magnetic field which is mediating transfer of energy and variability from the convective layer to the surface; one could also imagine that internal gravity waves are involved. Simulations of the generation and propagation of such gravity waves were presented by Cantiello et al. (2011).

The massive stars displaying strong large-scale fields also display variability and X-ray emission, implying that they also have activity originating in these subsurface convection layers. Exactly how a fossil field should co-exist with a convective dynamo is not clear. A very strong fossil field should inhibit the convection entirely; the field strength required though should be significantly above that corresponding to equipartition with the convection, and is probably not achieved in any star, at least for the iron-ionisation zone. A weaker field will certainly affect the operation of the dynamo and the way in which magnetic energy is transported upwards.

\section{Discussion}\label{disc}

With dynamo-generated fields, only one theory is required to explain their presence and nature. With fields which are a remnant from an earlier epoch, the questions of their nature and origin are separate. Above, I discussed the nature of fossil and failed fossil fields; I now speculate as to their origins: why do some stars host strong fossil fields, and why do most lack them? Any satisfying formation theory should also explain the range in fossil field strengths of more than two orders of magnitude and the variety of field geometries.

Since there is also a magnetic subset of HAeBe stars, fossil fields can be traced at least as far back as the pre-MS, if not further. The traditional hypothesis has been that variation in magnetic properties on the main sequence can be traced right back to the molecular cloud out of which the star was born. Indeed, the ISM does display a variation in magnetisation, albeit with much less than the range observed in stars. To explain the bimodality we could use some mechanism to destroy fields below a certain threshold; it is interesting to note that the lower cut-off of around 200 G corresponds to the strength at which the magnetic field is in equipartition with the gas pressure at the photosphere. This may or may not be a coincidence.

Another popular hypothesis has been that of mergers. A merger should produce such energy in differential rotation that it is sufficient to convert a fraction $\lesssim 10^{-5}$ of that energy into the magnetic energy of the eventual fossil equilibrium. Ferrario et al.\ (2009) point out that the merger product should be radiative and remain so until the main-sequence, in order to retain the field produced; as described in Braithwaite 2012, a convective star quickly loses any memory of its magnetic past. This might explain why the magnetic fraction increases towards greater masses (Power et al.\ 2007), since more massive pre-MS merger products are more likely to be and remain radiative.

An important clue as to the origin of fossil fields comes from binarity: the binary fraction amongst the magnetic stars has been found to be lower than in the non-magnetic stars (Abt \& Snowden 1973). There is apparently a complete lack of Ap stars in binaries with periods of less than about 3 days, except for one known example (HD 200405) with a period of 1.6 days. Obviously, a lack of magnetic stars in close binaries is what one would expect if they are the result of mergers. In the ISM variation hypothesis, it would also not be surprising: a more strongly magnetised cloud core loses more angular momentum and is less likely to break up and form a binary. Puzzling though would still be the binaries with one Ap and one normal A star.

Whatever the process to produce fossil fields, the issue of helicity generation is crucial. Magnetic helicity, a scalar quantity defined as $H\equiv(1/8\pi)\int \!{\bf A}\cdot{\bf B}\;{\rm d}V$ where ${\bf A}$ is the vector potential (${\bf B}=\bm{\nabla}\times{\bf A}$), is what ultimately determines the strength of the fossil equilibrium. This is because it is approximately conserved during relaxation to equilibrium, so that the equilibrium magnetic energy, which one can express in terms of a mean magnetic field strength $\bar B$ as $E_{\rm eq}=(4\pi/3)R^3\bar B^2/8\pi$, is given by $E_{\rm eq}=\zeta H/R$ where $R$ is the stellar radius and $\zeta$ is a factor of order unity (either positive or negative) which depends on the geometry of the equilibrium. According to this picture, stars with fossil fields were given a lot of helicity from some process, so that their pre-determined equilibrium field strengths were high. A star with a subgauss field, on the other hand, contains very little helicity and has a correspondingly low pre-determined equilibrium field strength, so low that the equilibrium is never actually reached within the lifetime of the star, owing to the increase of the field evolution timescale $\tau_{\rm A}^2\Omega$ as the field becomes weaker. Since we expect at least the intermediate-mass stars to host a convective dynamo during their pre-MS, we are left to conclude that this dynamo generates little helicity. Thinking about negative or positive helicity as a net left-handed or right-handed twist, it seems that the pre-MS dynamo does not produce the required symmetry-breaking for a large non-zero helicity.

Finally, a comment regarding rotation. Whilst `non-magnetic' A stars generally have rotation periods of a few hours to a day, most Ap stars have periods between one and ten days, and some have periods much greater (see e.g. Abt \& Morrell 1995). The slowest periods measured are of order decades and in several cases there are only lower limits.  One can speculate that the slow rotation of magnetic stars is a consequence of their magnetism rather than the other way around, especially if the star becomes magnetic while there is still circumstellar material onto which excess angular momentum can be offloaded. Although star-disc interaction is poorly understood, it seems logical that a strong stellar magnetic field leads to a large disc truncation radius, resulting in the spindown of the star until the co-rotation radius becomes comparable. Any theory might struggle though to explain rotation periods of order a century. However, the slow rotation of Ap stars holds only in a very broad sense; rapidly rotating examples like CU Virginis (0.5 d) exist as well. Any given explanation for the slow average rotation may well miss the most important clue: the astonishingly large range in rotation periods, of more than four orders of magnitude.

\end{document}